\documentclass[prc,twocolumn,showpacs,preprintnumbers,amsmath,amssymb]{revtex4}
\usepackage{graphicx}% Include figure files
\usepackage{dcolumn}% Align table columns on decimal point
\usepackage{bm}% bold math

\begin{document}

%\preprint{}

\title{Shell-Model Calculations for
$_{\Lambda}^{16}$O and $_{\Lambda}^{17}$O Using\\
Microscopic Effective Interactions with $\Sigma$
Degrees of Freedom}% Force line breaks with \\

\author{Shinichiro Fujii}
 \email{sfujii@postman.riken.go.jp}
\affiliation{%
RI Beam Factory Project Office,
RIKEN (The Institute of Physical and Chemical Research),
Wako 351-0198, Japan
}%
\author{Ryoji Okamoto}
\author{Kenji Suzuki}
\affiliation{
Department of Physics, Kyushu Institute of Technology,
Kitakyushu 804-8550, Japan
}

\date{\today}% It is always \today, today,

\begin{abstract}
Shell-model calculations in a large model space are performed for
$^{16}_{\Lambda}$O and $^{17}_{\Lambda}$O.
Effective interactions with degrees of freedom of $\Sigma$
in addition to $\Lambda$ and nucleons are derived from hyperon-nucleon and
nucleon-nucleon interactions within the framework of
the unitary-model-operator approach.
Effects of the $\Sigma$ degrees of freedom and the parity-mixing intershell
coupling on the $\Lambda$ hypernuclear structure are investigated,
employing the Nijmegen NSC97a-f and NSC89 hyperon-nucleon potentials.

\end{abstract}

\maketitle

\section{\label{sec:Introduction}Introduction}

One of the challenging problems in theoretical studies of $\Lambda$
hypernuclei is to describe their properties, starting from
hyperon-nucleon (${\it YN}$) and nucleon-nucleon (${\it NN}$) interactions
given in free space.
The nuclear shell-model approach
would be one of the promising methods for this problem
over a wide range of mass numbers of $\Lambda$ hypernuclei.
In shell-model calculations, however,
we need to introduce an effective interaction
because of a limited dimension of a shell-model space.

Microscopic derivation of an effective interaction for nuclear shell-model
calculations is a fundamental problem
for a deeper understanding of nuclei.
The $G$-matrix has often been introduced as an approximate
effective interaction, taking into account the state dependence,
the medium effect, and the Pauli-blocking effect in a nucleus.
Hao {\it et al.} performed the shell-model calculation for
$_{\Lambda}^{16}$O, deriving the $\Lambda {\it N}$ $G$-matrix
for a finite system~\cite{Hao93}.
Afterward, Tzeng {\it et al.}
have developed their approach to the effective $\Lambda {\it N}$ interaction
by calculating core polarization diagrams and folded diagrams~\cite{Kuo90}
in addition to the bare $G$-matrix~\cite{Tzeng99}.
They have also proposed the two-frequency shell model by introducing
different frequencies of the harmonic-oscillator (h.o.) wave functions
in order to describe different spreads of the wave functions of
$\Lambda$ and nucleons~\cite{Tzeng00}.

As for the treatment of the $\Lambda$ wave function,
Motoba has discussed that the mixing of higher nodal h.o. basis
functions is needed for the description of the weekly-bound $0p$ states of
$\Lambda$ in the study of $_{\Lambda}^{16}$O together with
$_{\Lambda}^{17}$O~\cite{Motoba98,Dalitz97},
using the YNG interaction proposed by Band\=o and Yamamoto~\cite{YNG85}.

The $_{\Lambda}^{16}$O is a representative hypernucleus for which experimental
data are comparatively accumulated.
Furthermore, a high-resolution $\gamma$-ray spectroscopy experiment for
$_{\Lambda}^{16}$O has been performed at BNL~\cite{Tamura01}.
Useful information on fine structures reflecting the properties of
the underlying ${\it YN}$ interactions should be obtained.
In such a situation, it is of highly interest to examine to what extent
the ${\it YN}$ interactions proposed up to now reproduce experimental data on
the $\Lambda$ hypernuclear structure.
For this reason, the importance of accurate derivation of the effective
interaction for the shell-model calculation has been growing.

In our previous works~\cite{Fujii99,Fujii00},
we have proposed a method for a microscopic description of $\Lambda$
hypernuclei
within the framework of the unitary-model-operator approach
(UMOA)~\cite{Suzuki94}. 
The UMOA is a many-body theory that leads to
an energy-independent and Hermitian effective interaction which possesses
the decoupling property~\cite{Suzuki80} between two states
in a model space and an excluded one.
An effective Hamiltonian is given by a unitary transformation
of an original Hamiltonian.
We here note that this type of effective interaction
has been used in recent accurate calculations for light
nuclei, for example, the no-core shell model~\cite{Navratil00}
and the method of the effective-interaction hyperspherical
harmonics~\cite{Barnea01}.

We applied the UMOA to the calculations of $\Lambda$
single-particle energies in $^{17}_{\Lambda}$O and
$^{41}_{\Lambda}$Ca~\cite{Fujii00}, using
${\it YN}$ interactions given by the Nijmegen~\cite{NSC89,NSC97} and
the J\"ulich~\cite{Juelich94} groups.
Some reasonable results were obtained, such as the small spin-orbit
splitting of $\Lambda$~\cite{Ajimura01} compared with those in nuclei
though the calculated energies considerably depend
on the ${\it YN}$ interactions employed.
It was also confirmed that the mixing of higher nodal h.o.
basis states was important for the description of the $\Lambda$ wave function.

In this work, we try to perform shell-model calculations for $_{\Lambda}^{16}$O
in addition to $_{\Lambda}^{17}$O.
In the shell-model calculations for $\Lambda$ hypernuclei made so far,
the effects of the $\Sigma {\it N}$ channel have been treated as
renormalization into a $\Lambda {\it N}$ effective interaction in many cases.
The $\Sigma$ degrees of freedom have not been treated explicitly
in the shell-model calculations.  Therefore, it is interesting to derive
an effective ${\it YN}$ interaction which includes not only
the $\Lambda {\it N}$ channel but also the $\Sigma {\it N}$ one,
and to apply such an effective interaction
to shell-model calculations for $\Lambda$ hypernuclei.
Another difference of our approach from the usual shell-model calculations is
that we do not employ the experimental single-particle energies of nucleons.
Instead, we use the calculated single-particle energies of nucleons
which are determined with the effective ${\it NN}$ interactions.
We also use the single-particle energies of $\Lambda$ and $\Sigma$ determined
in a similar way.

This paper is organized as follows. In Sec.~II, the procedure
for the shell-model calculation is given.
The calculated results for $_{\Lambda}^{17}$O and $_{\Lambda}^{16}$O
using the Nijmegen soft-core 97 (NSC97)~\cite{NSC97} and NSC89~\cite{NSC89}
potentials are shown in Sec.~III.
Finally, concluding remarks are made in Sec.~IV.

\section{Calculation Method}

\subsection{Effective Hamiltonian}

We first consider a Hamiltonian of a hypernuclear system consisting of
nucleons and one $\Lambda$ (or $\Sigma$).
It may be written in the second-quantization form as
\begin{eqnarray}
\label{eq:Ham}
H&=&\sum_{\alpha \beta }
\langle \alpha |t_{\it N}|\beta \rangle c_{\alpha }^{\scriptscriptstyle \dag }
c_{\beta }
+\frac{1}{4}\sum_{\alpha \beta \gamma \delta }
\langle \alpha \beta |v_{{\it N}_{\rm 1}{\it N}_{\rm 2}}|\gamma \delta \rangle
c_{\alpha }^{\dag }c_{\beta }^{\dag }c_{\delta }c_{\gamma }\nonumber \\
&&+\sum_{\mu \nu }
%&&+\sum_{\stackrel {\scriptstyle \mu \nu }{{\it Y}=\Lambda, \Sigma}}
\langle \mu |t_{\it Y}+\Delta m|\nu \rangle 
d_{\mu }^{\dag }d_{\nu }\nonumber \\
&&+\sum_{\mu \alpha \nu \beta }
%&&+\sum_{\stackrel {\scriptstyle \mu \alpha \nu \beta }
%{{\it Y}=\Lambda ,\Sigma}}
\langle \mu \alpha |v_{\it YN}|\nu \beta \rangle
d_{\mu }^{\dag }c_{\alpha }^{\dag }c_{\beta }d_{\nu },
\end{eqnarray}
where $c^{\dag }$ ($c$) is the creation (annihilation)
operator for a nucleon
in the usual notation, and $d^{\dag }$ ($d$) is the creation
(annihilation)
operator for a hyperon, $\Lambda$ or $\Sigma$.
The kinetic energies of a nucleon and a hyperon are denoted by
$t_{\it N}$ and $t_{\it Y}$, respectively.
The quantities $v_{{\it N}_{\rm 1}{\it N}_{\rm 2}}$ and
$v_{\it YN}$
represent the ${\it NN}$ and ${\it YN}$ interactions, respectively.
The symbols $\alpha ,\beta, \gamma$, and $\delta$ are used for
the sets of quantum numbers of nucleon states,
and $\mu$ and $\nu$ for those of hyperon states.
The $|\alpha \beta \rangle $ is
the anti-symmetrized and normalized two-body ${\it NN}$ state.
The term $\Delta m$ denotes the difference between
the rest masses of $\Lambda$ and $\Sigma$.

In order to properly treat the short-range two-body correlation,
we introduce a unitary transformation of the Hamiltonian as
\begin{eqnarray}
\label{eq:eff_Ham}
\tilde{H}=e^{-S}He^{S},
\end{eqnarray}
where $S$ is the sum of anti-Hermitian two-body operators for
the ${\it NN}$ and ${\it YN}$ systems defined as 
\begin{eqnarray}
\label{eq:S}
S=S^{({\it NN})}+S^{({\it YN})},
\end{eqnarray}
with
\begin{eqnarray}
\label{eq:S_NN}
S^{({\it NN})}=\frac{1}{4}\sum_{\alpha \beta \gamma \delta}
\langle \alpha \beta |S_{{\it N}_{\rm 1}{\it N}_{\rm 2}}|\gamma \delta \rangle
c_{\alpha }^{\dag }c_{\beta }^{\dag }c_{\delta }c_{\gamma },
\end{eqnarray}
\begin{eqnarray}
\label{eq:S_YN}
S^{({\it YN})}=\sum_{\mu \alpha \nu \beta}
\langle \mu \alpha |S_{\it YN}|\nu \beta \rangle
d_{\mu }^{\dag }c_{\alpha }^{\dag }c_{\beta }d_{\nu }.
\end{eqnarray}

We adopt a cluster expansion of the unitarily transformed
Hamiltonian as
\begin{eqnarray}
\label{eq:cluster}
\tilde{H}=\tilde{H}^{(1)}+\tilde{H}^{(2)}+\cdot \cdot \cdot ,
\end{eqnarray}
where the first two terms are written explicitly as
\begin{eqnarray}
\label{eq:cluster_1}
\tilde{H}^{(1)}=\sum_{\alpha \beta }
\langle \alpha |h_{\it N}|\beta \rangle 
c_{\alpha }^{\dag }c_{\beta }
+\sum_{\mu \nu }
%+\sum_{\stackrel {\scriptstyle \mu \nu }{{\it Y}=\Lambda, \Sigma}}
\langle \mu |h_{\it Y}|\nu \rangle d_{\mu }^{\dag }d_{\nu },
\end{eqnarray}
%---------------
\begin{eqnarray}
\label{eq:cluster_2}
\tilde{H}^{(2)}&=&\frac{1}{4}
\sum_{\alpha \beta \gamma \delta}
\langle \alpha \beta |\tilde{v}_{{\it N}_{\rm 1}{\it N}_{\rm 2}}|
\gamma \delta \rangle
c_{\alpha }^{\dag }c_{\beta }^{\dag }c_{\delta }c_{\gamma }\nonumber\\
&&-\sum_{\alpha \beta }
\langle \alpha |u_{\it N}|\beta \rangle 
c_{\alpha }^{\dag }c_{\beta }\nonumber \\
&&+\sum_{\mu \alpha \nu \beta }
%&&+\sum_{\stackrel {\scriptstyle \mu \alpha \nu \beta }
%{{\it Y}=\Lambda, \Sigma}}
\langle \mu \alpha |\tilde{v}_{\it YN}|\nu \beta \rangle
d_{\mu }^{\dag }c_{\alpha }^{\dag }c_{\beta }d_{\nu }\nonumber\\
&&-\sum_{\mu \nu }
%&&-\sum_{\stackrel {\scriptstyle \mu \nu }{{\it Y}=\Lambda, \Sigma}}
\langle \mu |u_{\it Y}|\nu \rangle d_{\mu }^{\dag }d_{\nu }.
\end{eqnarray}
%-----------
Since the exponent $S$ in Eq.~(\ref{eq:eff_Ham}) is a two-body operator,
the one-body parts in $\tilde{H}$
are unchanged and given by
\begin{eqnarray}
\label{eq:h_N}
h_{\it N}=t_{\it N}+u_{\it N},
\end{eqnarray}
%--------------
\begin{eqnarray}
\label{eq:h_Y}
h_{\it Y}=t_{\it Y}+u_{\it Y}+\Delta m_{\it Y}.
\end{eqnarray}
%-------------
The terms $u_{k}$ for $k={\it N}$ and ${\it Y}$ are
the auxiliary single-particle potentials of a nucleon and a hyperon,
respectively, and at this stage they are arbitrary.
The quantities $\tilde{v}_{kl}$ for
$\{kl\}=\{ {\it N}_{\rm 1}{\it N}_{\rm 2}\} $ and
$\{ {\it YN}\}$ are the transformed two-body interactions for
the ${\it NN}$ and ${\it YN}$ systems, and they are given by
%--------------
\begin{eqnarray}
\label{eq:v_NN}
\tilde{v}_{{\it N}_{\rm 1}{\it N}_{\rm 2}}=
&&e^{-S_{{\it N}_{\rm 1}{\it N}_{\rm 2}}}
(h_{{\it N}_{\rm 1}}+h_{{\it N}_{\rm 2}}+v_{{\it N}_{\rm 1}N_{\rm 2}})
e^{S_{{\it N}_{\rm 1}N_{\rm 2}}}\nonumber\\
&&-(h_{{\it N}_{\rm 1}}+h_{{\it N}_{\rm 2}}),
\end{eqnarray}
%---------------
\begin{eqnarray}
\label{eq:v_YN}
\tilde{v}_{\it YN}=&&e^{-S_{\it YN}}
(h_{\it Y}+h_{\it N}+v_{\it YN})e^{S_{\it YN}}\nonumber\\
&&-(h_{\it Y}+h_{\it N}).
\end{eqnarray}
%------------
In nuclear many-body problems, it is important how to choose
the auxiliary potential $u_{k}$.
We introduce $u_{\it N}$ and $u_{\it Y}$ as self-consistent potentials
defined with the transformed two-body interactions
$\tilde{v}_{{\it N}_{\rm 1}{\it N}_{\rm 2}}$ and
$\tilde{v}_{\it YN}$ as
\begin{eqnarray}
\label{eq:u_N}
\langle \alpha |u_{\it N}| \beta \rangle
=\sum_{\xi \leq \rho _{\rm F}}
\langle \alpha \xi |\tilde{v}_{{\it N}_{\rm 1}{\it N}_{\rm 2}}|
\beta \xi \rangle ,
\end{eqnarray}
\begin{eqnarray}
\label{eq:u_Y}
\langle \mu |u_{\it Y}| \nu \rangle =\sum_{\xi \leq \rho _{\rm F}}
\langle \mu \xi |\tilde{v}_{\it YN}|\nu \xi \rangle,
\end{eqnarray}
where $\rho _{\rm F}$ is the uppermost occupied level, and the symbol
$\xi$ indicates an occupied state for nucleons.
If $u_{\it N}$ and $u_{\it Y}$ are defined as given in
Eqs.~(\ref{eq:u_N}) and (\ref{eq:u_Y}), it can be proved
that the second and fourth terms on the right-hand side of
Eq.~(\ref{eq:cluster_2}) are canceled by
the contributions of the bubble diagrams of the first and third terms,
respectively.
Therefore, the cluster terms $\tilde{H}^{(1)}$ and $\tilde{H}^{(2)}$ 
include only the one- and two-body operators,
respectively,
if we write them in the normal-product form with respect to particles
and holes~\cite{Suzuki94}.

The transformed Hamiltonian $\tilde{H}$ contains, in general,
three-or-more-body transformed interactions, even if the starting Hamiltonian
$H$ in Eq.~(\ref{eq:Ham}) does not include three-or-more-body interactions.
In the previous paper~\cite{Fujii00}, a method of evaluating
the three-body cluster (TBC) effect
has been presented for the calculation of
$\Lambda$ single-particle energies.
The TBC terms
are generated as the transformed three-body interactions
among the ${\it YNN}$ and ${\it NNN}$ systems,
and they contain generally two factors of the correlation operator $S_{kl}$
for \{$kl$\} $=$ \{${\it N}_{\rm 1}{\it N}_{\rm 2}$\} and \{${\it YN}$\}.
It has been verified that the matrix elements of
the correlation operator are quite small, and thus the TBC
contributions to the $\Lambda$ single-particle energies are
considerably smaller than the contributions from
the one- and two-body cluster terms.
The TBC contributions to the $\Lambda$ single-particle energies in
$_{\Lambda}^{17}$O were found to be at most 4\%
of the $\Lambda$ potential energy.
Therefore, we assume that the TBC terms do not have significant effects
on the energy levels in $_{\Lambda}^{16}$O, and the three-or-more-body
effective interactions are not included in the present calculation.

%------------------------------------------
\subsection{\label{sec:Mspace}Model space of
two-body ${\it NN}$ and ${\it YN}$ states}
%-----------------------------------------

An important problem in the present approach is how to determine the
two-body correlation operators
$S_{kl}$ for \{$kl$\} $=$ \{${\it N}_{\rm 1}{\it N}_{\rm 2}$\}
and \{${\it YN}$\} in Eqs.~\eqref{eq:S_NN} and \eqref{eq:S_YN}, respectively.
These operators are given as solutions to the equation of decoupling as
\begin{eqnarray}
\label{eq:decoupling}
Q_{kl}e^{-S_{kl}}
(h_{k}+h_{l}
+v_{kl})e^{S_{kl}}P_{kl}=0,
\end{eqnarray}
where $P_{kl}$ and $Q_{kl}$
%for \{$kl$\} $=$
%\{${{\it N}_{\rm 1}{\it N}_{\rm 2}}$\} and \{${\it YN}$\}
are projection operators which act in the space of two-body states and
project a two-body state onto the low-momentum model space
and the high-momentum space, respectively.
It has been proved that Eq.~(\ref{eq:decoupling})
for $S_{kl}$ can be solved in a nonperturbative way~\cite{Suzuki94,Fujii99}
under the conditions
\begin{eqnarray}
\label{eq:restriction}
P_{kl}S_{kl}P_{kl}=Q_{kl}S_{kl}Q_{kl}=0.
\end{eqnarray}
%-------------

\begin{figure}[t]
\includegraphics[height=.205\textheight]{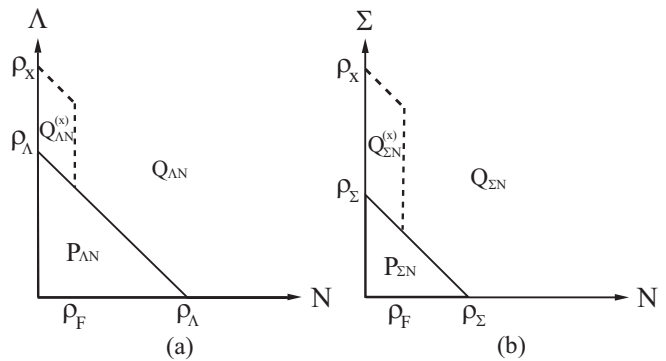}
\caption{\label{fig:1} The model space of the $\Lambda {\it N}$ (a)
and $\Sigma {\it N}$ (b) channels and its complement.}
\end{figure}

In order to actually calculate the effective interaction,
we need two-body model spaces
for the ${\it NN}$ and ${\it YN}$ channels.
We choose the same model space for the ${\it NN}$ channel
as in Ref.~\cite{Suzuki94}.
We here define the model space of the ${\it YN}$ channel as follows:
Two-body ${\it YN}$ states for ${\it Y}=\Lambda$ and $\Sigma$
are given by the product of the h.o. wave functions as
\begin{equation}
\label{eq:YNstate}
|\mu _{\it Y}\alpha _{\it N}\rangle = |n_{\it Y}l_{\it Y}j_{\it Y}m_{\it Y},
n_{\it N}l_{\it N}j_{\it N}m_{\it N}\rangle
\end{equation}
with the h.o. quantum numbers,
$\{n_{\it Y},l_{\it Y},j_{\it Y},m_{\it Y}\}$
and $\{n_{\it N},l_{\it N},j_{\it N},m_{\it N}\}$, of a hyperon and a nucleon,
respectively.
The model space of the ${\it YN}$ channel
$P_{\Lambda {\it N}}+P_{\Sigma {\it N}}$
and its complement $Q_{\Lambda {\it N}}+Q_{\Sigma {\it N}}$
are defined with boundary numbers $\rho _{\Lambda}$ and $\rho _{\Sigma}$ as
\begin{eqnarray}
\label{eq:PQLam}
|\mu _{\Lambda}\alpha _{\it N}\rangle 
\in \left\{ 
\begin{array}{ll}
\displaystyle {P_{\Lambda {\it N}}} &
{\rm if \ } 2n_{\Lambda}+l_{\Lambda}
+2n_{\it N}+l_{\it N} \leq \rho _{\Lambda }, \\
\displaystyle {Q_{\Lambda {\it N}}} & {\rm otherwise}\\
\end{array}\right.
\end{eqnarray}
and
\begin{eqnarray}
\label{eq:PQSig}
|\mu _{\Sigma}\alpha _{\it N}\rangle 
\in \left\{ 
\begin{array}{ll}
\displaystyle {P_{\Sigma {\it N}}} &
{\rm if \ } 2n_{\Sigma}+l_{\Sigma}
+2n_{\it N}+l_{\it N} \leq \rho _{\Sigma }, \\
\displaystyle {Q_{\Sigma {\it N}}} & {\rm otherwise}.\\
\end{array}\right.
\end{eqnarray}
Note that the numbers $\rho _{\Lambda}$ and $\rho _{\Sigma}$
are zero or positive integers.
In Fig.~\ref{fig:1}, the model space and its complement are shown.
The ${\it YN}$ states in the space
$Q_{\Lambda {\it N}}^{\rm (X)}+Q_{\Sigma {\it N}}^{\rm (X)}$
specified by the numbers $\rho _{\Lambda}, \rho _{\Sigma}, \rho _{\rm F}$,
and $\rho _{\rm X}$ in Fig.~\ref{fig:1} should be excluded due to the Pauli
principle for nucleons, and defined as
\begin{eqnarray}
\label{eq:QLNx}
|\mu _{\Lambda}\alpha _{\it N}\rangle 
\in Q_{\Lambda {\it N}}^{\rm (X)}\ \ 
&&\ {\rm if \ } \rho _{\Lambda} < 
 2n_{\Lambda}+l_{\Lambda}
+2n_{\it N}+l_{\it N} 
 \leq \rho _{\rm X} \nonumber \\
&&{\rm and \ } 0 \leq 2n_{\it N}
+l_{\it N} 
\leq \rho _{\rm F}
\end{eqnarray}
and
\begin{eqnarray}
\label{eq:QSNx}
|\mu _{\Sigma}\alpha _{\it N}\rangle 
\in Q_{\Sigma {\it N}}^{\rm (X)}\ \ 
&&\ {\rm if \ } \rho _{\Sigma} < 
 2n_{\Sigma}+l_{\Sigma}
+2n_{\it N}+l_{\it N} 
 \leq \rho _{\rm X} \nonumber \\
&&{\rm and \ } 0 \leq 2n_{\it N}
+l_{\it N} 
\leq \rho _{\rm F}.
\end{eqnarray}
The number $\rho _{\rm F}$ denotes the highest occupied orbit in the core
nucleus and is taken as $\rho _{\rm F}=1$ in the present case of $^{16}$O.
The value of $\rho _{\rm X}$ should be chosen as large as possible
so as to exclude the ${\it YN}$ states
in the Pauli-blocked $Q_{\it YN}^{\rm (X)}$ space.
In the present calculation, we take as $\rho _{\rm X}=12$.
The values of $\rho _{\Lambda}$ and $\rho _{\Sigma}$, in principle,
should be taken as a large value 
so that the results become independent of $\rho _{\Lambda}$ and 
$\rho _{\Sigma}$.
As for $\rho _{\Lambda}$, we take as $\rho _{\Lambda}=8$
which has been shown to be sufficiently large in
the previous work~\cite{Fujii99}.
The $\rho _{\Sigma}$-dependence of calculated energy levels 
in $_{\Lambda}^{17}$O and $_{\Lambda}^{16}$O will be discussed in detail
in Sec.~III.

The effective interaction $\tilde{v}_{\it YN}$
in Eq.~\eqref{eq:v_YN} is determined by solving the decoupling equation
in Eq.~\eqref{eq:decoupling}
between the model space
$P_{\it YN}=P_{\Lambda {\it N}}+P_{\Sigma {\it N}}$ and the space
$Q_{\it YN}=(Q_{\Lambda {\it N}}-Q_{\Lambda {\it N}}^{\rm (X)})
+(Q_{\Sigma {\it N}}-Q_{\Sigma {\it N}}^{\rm (X)})$.
The detailed procedure for solving the decoupling equation has been given
in Ref.~\cite{Fujii99}.

\subsection{\label{sec:Diagonalization}Shell-model diagonalization}

We proceed to discuss the calculation procedure for
the shell-model diagonalization. 
The shell-model spaces we adopt are given by
%--------------
\begin{equation}
\label{eq:space17}
d^{\dagger}_{\Lambda}|\phi _{0}\rangle \oplus
d^{\dagger}_{\Lambda} a^{\dagger}b^{\dagger}|\phi _{0}\rangle \oplus
d^{\dagger}_{\Sigma} a^{\dagger}b^{\dagger}|\phi _{0}\rangle
\end{equation}
%----------
for $_{\Lambda}^{17}$O and
\begin{equation}
\label{eq:space16}
d^{\dagger}_{\Lambda}b^{\dagger}|\phi _{0}\rangle \oplus
d^{\dagger}_{\Sigma}b^{\dagger}|\phi _{0}\rangle \oplus
d^{\dagger}_{\Lambda} a^{\dagger}b^{\dagger}b^{\dagger}|\phi _{0}\rangle
\oplus
d^{\dagger}_{\Sigma} a^{\dagger}b^{\dagger}b^{\dagger}|\phi _{0}\rangle
\end{equation}
%-------------
for $_{\Lambda}^{16}$O,
where $d^{\dagger}_{\Lambda}$ ($d^{\dagger}_{\Sigma}$) is the creation operator
of a $\Lambda$ ($\Sigma$),
and $a^{\dagger}$ and $b^{\dagger}$ are
the creation operators of a particle and a hole, respectively, for nucleons.
The state $|\phi _{0}\rangle$ is the unperturbed ground state of
the core nucleus which is the particle-hole vacuum
satisfying $a|\phi _{0}\rangle = b|\phi _{0}\rangle=0$.

In general, the transformed Hamiltonian
$\tilde{H}$ in Eq.~\eqref{eq:eff_Ham} includes three-or-more-body
effective interactions.
In the present calculation, as mentioned before,
we neglect the many-body effective
interactions, and take the one- and two-body parts in $\tilde{H}$.
In this approximation, the direct coupling of
$d^{\dagger}_{\Lambda}|\phi _{0}\rangle$ and
$d^{\dagger}_{\Lambda}a^{\dagger}a^{\dagger}b^{\dagger}b^{\dagger}
|\phi_{0}\rangle$ in $_{\Lambda}^{17}$O does not occur anymore.
This is because the operator $S_{{\it N}_{\rm 1}{\it N}_{\rm 2}}$
in Eq.~\eqref{eq:S_NN} is determined so that the transformed
Hamiltonian does not contain interactions inducing
two-particle-two-hole ($2p$-$2h$) excitations
in the ground state of the core nucleus~\cite{Suzuki94}.
The same discussion applies to 
the direct coupling of
$d^{\dagger}_{\Lambda}b^{\dagger}|\phi _{0}\rangle$ and
$d^{\dagger}_{\Lambda}a^{\dagger}a^{\dagger}b^{\dagger}b^{\dagger}b^{\dagger}
|\phi_{0}\rangle$ in $_{\Lambda}^{16}$O.
On these considerations, we take only the shell-model spaces as in
Eqs.~\eqref{eq:space17} and \eqref{eq:space16}
in which the effective Hamiltonian is diagonalized.

It should be noted that since we diagonalize the unitarily
transformed Hamiltonian $\tilde{H}$
in Eq.~\eqref{eq:eff_Ham}, a true eigenstate of the original Hamiltonian
$H$ can be given by a transformed state.
That is to say, an eigenstate of $H$ denoted by $|\Psi _{k}\rangle$ is given by
$e^{S}|\psi _{k}\rangle$, where $|\psi _{k}\rangle$ is an eigenstate of
the transformed Hamiltonian $\tilde{H}$.
The correlated ground state of
the core nucleus $|\Phi _{0}\rangle$ is related
to the unperturbed shell-model ground state
$|\phi _{0}\rangle$ as $|\Phi_{0} \rangle =e^{S} |\phi_0\rangle$.
In general, $|\Phi_{0}\rangle$ contains $2p$-$2h$,
$4p$-$4h$, and higher-order particle-hole components through
the unitary transformation $e^{S}$ with the two-body correlation operator $S$.
In a similar way, the transformed state $|\Psi _{k}\rangle$ contains
many-particle-many-hole components consistently with the correlations
in the ground state of the core nucleus.

We here want to discuss the $\Lambda$-$\Sigma$ coupling three-body force
of which effect has been pointed out
by Tzeng {\it et al}.~\cite{Tzeng99,Tzeng00}.
In the present shell-model calculation we neglect the transformed
three-or-more-body interactions, but this does not mean to neglect the
$\Lambda$-$\Sigma$ three-body force.
In our approach the $\Sigma {\it N}$-$\Lambda {\it N}$ coupling terms remain
in the transformed Hamiltonian,
if we include the $\Sigma {\it N}$ states in the ${\it YN}$ model space.
Therefore, the $\Sigma {\it N}$-$\Lambda {\it N}$ coupling is evaluated as
configuration mixing of $\Sigma$-nucleons states into $\Lambda$-nucleons ones.

In shell-model calculations,
spurious states caused by the center-of-mass (c.m.) motion often mix
with physical states.
In the present case,
the $1p$-$1h$ spurious
$1^{-}(T=0)$ state in $^{16}$O affects low-lying physical states, especially,
the $3/2^{-}(T=0)$ and $1/2^{-}(T=0)$ states
in $_{\Lambda}^{17}$O~\cite{Yamada85}.
In order to remove the spurious c.m. state, we add the following term
\begin{eqnarray}
\label{eq:H_b}
H_{\beta}&=&
\beta |1^{-}\, _{\rm c.m.}\rangle
\langle1^{-}\, _{\rm c.m.}|
\end{eqnarray}
to the effective Hamiltonian, and then the Hamiltonian is diagonalized.
We take as $\beta=3\hbar\omega$ in Eq.~\eqref{eq:H_b}
with the h.o. frequency $\hbar\omega$,
and we eliminate the spurious $1^{-}$ state from the low-lying states
under consideration.

As for the value of $\hbar\omega$, we take as $\hbar\omega=14$MeV
because the result tends to the saturation minimum of the binding energy
in $^{16}$O at close this value~\cite{Suzuki86}.
We employ the same value $14$MeV for $\hbar\omega$ of the hyperons $\Lambda$
and $\Sigma$.
In general, the spreads of the wave functions of the hyperons and nucleons
are different from each other.
If one tries to describe the states of the hyperons and nucleons
using the h.o. wave functions, one may choose the 
different frequencies as done in the two-frequency
shell model~\cite{Tzeng00}.
In the present work, however, we take the values of $\hbar \omega$ commonly
for the hyperons and nucleons,
because the final result in the shell-model calculation should be,
in principle, independent of $\hbar\omega$
if we take a sufficiently large model space 
in the calculation.

Since we diagonalize the effective Hamiltonian
in the space of the particle-hole states,
we should remove unlinked-diagram contributions in a suitable way.
In our approach, non-diagonal matrix elements of the one-body
part of the nucleon remain in the effective Hamiltonian, which induces
the $1p$-$1h$ excitation and causes the unlinked-diagram effect.
In order to remove the unlinked terms,
we calculate separately
the correlation energy $E_{\rm c}$ of the core nucleus in the space of
$|\phi _{0}\rangle \oplus a^{\dagger}b^{\dagger}|\phi _{0}\rangle$.
We then subtract $E_{\rm c}$ from the eigenvalue $E_{\rm SM}$
of the shell-model effective Hamiltonian.
In the case of $_{\Lambda}^{17}$O,
the value of $E_{\rm SM}-E_{\rm c}$ corresponds to the binding energy of
$\Lambda$ measured from the $^{16}$O+$\Lambda$
threshold.

\section{\label{sec:Results}Results and Discussion}

We performed calculations employing 
the Nijmegen soft-core 97 (NSC97)~\cite{NSC97} and NSC89~\cite{NSC89}
potentials for the ${\it YN}$ interaction.
As for the ${\it NN}$ interaction,
we choose the Paris~\cite{Paris80} potential.
All the interaction matrix elements of the Hamiltonian
in Eq.~\eqref{eq:eff_Ham} are derived from these bare interactions
within the framework of the UMOA.
In these calculations we do not introduce any adjustable parameters
and experimental values such as single-particle energies of
$\Lambda$, $\Sigma$, and nucleons.
This sort of microscopic calculation would be
worthy of revealing the states of the present ${\it YN}$ interactions.

\subsection{\label{subsec:O-17-L}Structure of $_{\Lambda}^{17}$O}

\begin{figure}[b]
\includegraphics[height=.25\textheight]{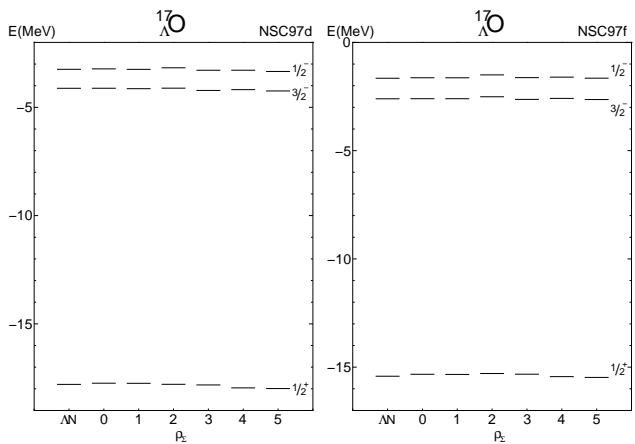}
\caption{\label{fig:2} The calculated energy levels in $_{\Lambda}^{17}$O
for the NSC97d and NSC97f potentials as a function of $\rho _{\Sigma}$.
The energy levels for ``$\Lambda {\it N}$" are the results for the case without
the $\Sigma {\it N}$ channel in the model space.}
\end{figure}

\begin{table*}[t]
\caption{\label{tab:table1} The values of the calculated energy levels in
$_{\Lambda}^{17}$O for the NSC97a-f potentials for $\rho _{\Sigma}=5$.
The quantity $\Delta E_{ls}$ stands for the magnitude of
the $\Lambda$ spin-orbit splitting defined as
$\Delta E_{ls}=E(1/2^{-})-E(3/2^{-})$.
The values in parentheses denote the results using
the renormalized ``$\Lambda {\it N}$" effective interaction.
 All energies are in MeV.}
\begin{ruledtabular}
    \begin{tabular}{ccccccc}
                      &   NSC97a  &   NSC97b   &   NSC97c   &  NSC97d   &  NSC97e   &  NSC97f   \\ \hline
 $1/2^{-}$            &  $ -3.29$ &  $ -3.19$  &   $ -3.51$ & $ -3.24$  & $ -2.82$  & $ -1.65$  \\
                      &  $(-3.44)$&  $(-3.37)$ &  $(-3.67)$ & $(-3.34)$ & $(-2.87)$ & $(-1.65)$ \\
 $3/2^{-}$            &  $ -3.72$ &  $ -3.75$  &   $ -4.23$ & $ -4.12$  & $ -3.77$  & $ -2.60$  \\
                      &  $(-3.91)$&  $(-3.95)$ &  $(-4.42)$ & $(-4.24)$ & $(-3.84)$ & $(-2.64)$ \\
 $1/2^{+}$            &  $-17.04$ &  $-17.14$  &   $-17.91$ & $-17.79$  & $-17.26$  & $-15.42$  \\
                      & $(-17.39)$&  $(-17.46)$& $(-18.23)$ & $(-17.99)$& $(-17.40)$& $(-15.47)$\\
 $\Delta E_{ls}$      &  $  0.44$ &  $  0.56$  &   $  0.72$ & $  0.88$  & $  0.94$  & $  0.95$  \\
                      &  $(0.47)$ &  $ (0.58)$ &   $(0.74)$ & $ (0.90)$ & $ (0.97)$ & $ (0.99)$ \\
    \end{tabular}
\end{ruledtabular}
\end{table*}

In Fig.~\ref{fig:2}, we first show
the calculated energy levels in $_{\Lambda}^{17}$O for the NSC97d and NSC97f
potentials as a function of $\rho _{\Sigma}$.
The results correspond to the
$\Lambda$ single-particle energies including the effect of core polarization.
One can see that the results for both potentials are stable for the change
of the values of $\rho _{\Sigma}$, and almost the same as the results
for ``$\Lambda {\it N}$".
The ``$\Lambda {\it N}$" means that the $\Sigma {\it N}$ channel
is not included in the model space.
This suggests that the effects of the $\Sigma {\it N}$ channel into
the $\Lambda {\it N}$ effective interaction can be well renormalized.
It has been confirmed that
this tendency of the convergence is also observed for the other NSC97 models.

In Table~\ref{tab:table1}, we tabulate the calculated energy levels
in $_{\Lambda}^{17}$O with the values of
the $\Lambda$ spin-orbit splitting for the NSC97a-f potentials.
In this table the values for $\rho _{\Sigma}=5$ are presented
as the sets of convergent results in this study,
and we also list the values in parentheses
which are the results for ``$\Lambda {\it N}$" for reference.
The results show that the energies for the NSC97c are the most attractive
in the NSC97 models,
on the other hand, those for the NSC97f are the least attractive.
This trend is also seen in the calculation for nuclear matter as in
Ref.~\cite{NSC97}.
We also see that the $\Lambda$ spin-orbit splittings become larger
from the NSC97a to NSC97f.
Recently, the magnitude of the $\Lambda$ spin-orbit splitting in
$_{\Lambda}^{13}$C has been established experimentally as
$\Delta E_{ls}(_{\Lambda}^{13}{\rm C})=E(1/2^{-})-E(3/2^{-})
=152\pm 54({\rm stat})\pm 36({\rm syst})$
keV~\cite{Ajimura01}.
Our results of the $\Lambda$ spin-orbit splitting in $_{\Lambda}^{17}$O
for the NSC97 models may considerably larger than the value
suggested from the experimental result of $_{\Lambda}^{13}$C.

We note here that the present results in Table~\ref{tab:table1}
agree well with those obtained in the previous work~\cite{Fujii00}
in which the calculation was made perturbatively.
We may say that both methods, the shell-model diagonalization and
the perturbative method, are workable in the calculation of
the $\Lambda$ single-particle energies in $\Lambda$ hypernuclei
which have the simple structure.
In the following subsection,
we shall proceed to study a more complex system, namely, $_{\Lambda}^{16}$O
by the shell-model diagonalization.

\subsection{\label{subsec:O-16-L}Structure of $_{\Lambda}^{16}$O}

Experimental energy levels in $_{\Lambda}^{16}$O are usually relative to
the ground state of $^{15}$O.
Since we employ the particle-hole formalism,
the results of the shell-model diagonalization
are relative to the ground state of $^{16}$O.
Therefore, we subtract the mass difference between $^{15}$O and $^{16}$O
from the calculated results in order to compare our results
with the experimental spectrum.
The mass difference is computed by the shell-model diagonalization
in the space of the $1h$+($1p$-$2h$) configuration, using the nucleon parts of
the effective Hamiltonian in Eq.~\eqref{eq:eff_Ham}.

\begin{figure}[b]
\includegraphics[height=.25\textheight]{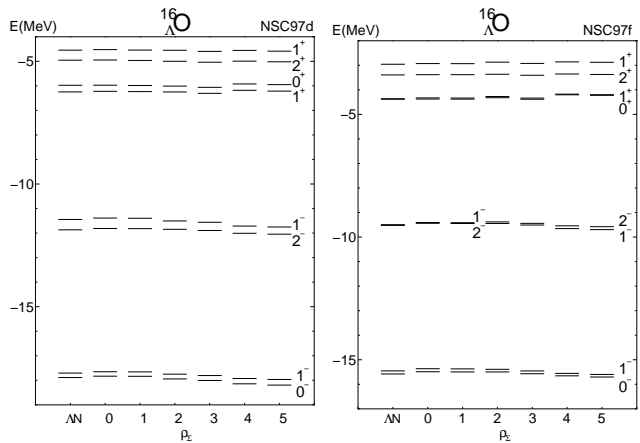}
\caption{\label{fig:3} The $\rho _{\Sigma}$-dependence of the calculated
results of low-lying states in $_{\Lambda}^{16}$O.}
\end{figure}

\begin{table*}[t]
\caption{\label{tab:table2} The calculated energy values of the low-lying
states in $_{\Lambda}^{16}$O 
for the NSC97a-f potentials for $\rho _{\Sigma}=5$.
The quantity $\Delta E_{1,2}$
stands for the magnitude of the splitting between the $1_{2}^{-}$ and
$2_{1}^{-}$ states defined as $\Delta E_{1,2}=E(1_{2}^{-})-E(2_{1}^{-})$,
and $\Delta E_{1,0}$ is defined as $\Delta E_{1,0}=E(1_{1}^{-})-E(0_{1}^{-})$.
The values in parentheses denote the results using
the renormalized ``$\Lambda {\it N}$" effective interaction.
 All energies are in MeV.}
\begin{ruledtabular}
    \begin{tabular}{ccccccc}
                       &   NSC97a   &   NSC97b   &   NSC97c   &   NSC97d   &  NSC97e   &  NSC97f \\ \hline
 $1_{2}^{+}$           & $  -4.38 $ & $  -4.38 $ & $  -4.80 $ & $  -4.59 $ & $  -4.15 $ & $  -2.88 $ \\
                       & $( -4.26)$ & $( -4.25)$ & $( -4.69)$ & $( -4.55)$ & $( -4.17)$ & $( -2.96)$ \\
 $2_{1}^{+}$           & $  -4.50 $ & $  -4.58 $ & $  -5.13 $ & $  -5.02 $ & $  -4.63 $ & $  -3.37 $ \\
                       & $( -4.36)$ & $( -4.43)$ & $( -4.99)$ & $( -4.95)$ & $( -4.61)$ & $( -3.39)$ \\
 $0_{1}^{+}$           & $  -4.98 $ & $  -5.16 $ & $  -5.90 $ & $  -5.95 $ & $  -5.63 $ & $  -4.22 $ \\
                       & $( -4.86)$ & $( -5.04)$ & $( -5.80)$ & $( -5.98)$ & $( -5.72)$ & $( -4.38)$ \\
 $1_{1}^{+}$           & $  -5.67 $ & $  -5.76 $ & $  -6.38 $ & $  -6.21 $ & $  -5.74 $ & $  -4.20 $ \\
                       & $( -5.60)$ & $( -5.70)$ & $( -6.32)$ & $( -6.25)$ & $( -5.83)$ & $( -4.36)$ \\
 $1_{2}^{-}$           & $ -10.74 $ & $ -10.93 $ & $ -11.77 $ & $ -11.75 $ & $ -11.33 $ & $  -9.70 $ \\
                       & $(-10.25)$ & $(-10.45)$ & $(-11.32)$ & $(-11.44)$ & $(-11.08)$ & $( -9.50)$ \\
 $2_{1}^{-}$           & $ -11.69 $ & $ -11.71 $ & $ -12.36 $ & $ -12.04 $ & $ -11.43 $ & $  -9.58 $ \\
                       & $(-11.39)$ & $(-11.43)$ & $(-12.08)$ & $(-11.87)$ & $(-11.31)$ & $( -9.53)$ \\
 $1_{1}^{-}$           & $ -17.24 $ & $ -17.34 $ & $ -18.14 $ & $ -17.96 $ & $ -17.43 $ & $ -15.60 $ \\
                       & $(-16.85)$ & $(-16.96)$ & $(-17.76)$ & $(-17.71)$ & $(-17.23)$ & $(-15.45)$ \\
 $0_{1}^{-}$           & $ -17.93 $ & $ -17.93 $ & $ -18.57 $ & $ -18.20 $ & $ -17.57 $ & $ -15.70 $ \\
                       & $(-17.48)$ & $(-17.51)$ & $(-18.12)$ & $(-17.89)$ & $(-17.32)$ & $(-15.58)$ \\
 $\Delta E_{1,2}$      & $   0.96 $ & $   0.78 $ & $   0.60 $ & $   0.29 $ & $   0.10 $ & $  -0.12 $ \\
                       & $  (1.15)$ & $  (0.98)$ & $  (0.75)$ & $  (0.43)$ & $  (0.22)$ & $  (0.03)$ \\
 $\Delta E_{1,0}$      & $   0.69 $ & $   0.59 $ & $   0.41 $ & $   0.23 $ & $   0.13 $ & $   0.11 $ \\
                       & $  (0.64)$ & $  (0.56)$ & $  (0.36)$ & $  (0.18)$ & $  (0.10)$ & $  (0.13)$ \\
    \end{tabular}
\end{ruledtabular}
\end{table*}

In Fig.~\ref{fig:3}, we show the $\rho _{\Sigma}$-dependence of the calculated
energies for low-lying states in $_{\Lambda}^{16}$O for
the NSC97d and NSC97f.
One can see that
the energy levels of the negative parity states become slightly more attractive
as the value of $\rho _{\Sigma}$ becomes larger.
We may say, however, that the splittings of the ground-stated doublet
($0_{1}^{-},1_{1}^{-}$) hardly change, and thus almost convergent results
are obtained.
On the other hand, in the first-excited doublet ($1_{2}^{-},2_{1}^{-}$),
the splittings become slightly smaller as $\rho _{\Sigma}$ increases.
These trends have also been observed in the results for the other NSC97 models.

There are some arguments that splittings of the spin doublets
($J_{>,<}=J_{\rm core} \pm s^{\Lambda}_{1/2}$) in $\Lambda$ hypernuclei
depend on the spin-dependent $\Lambda {\it N}$ interactions, such as
the spin-spin and tensor interactions~\cite{Millener85,NSC97}.
In addition, the $\Sigma {\it N}$-$\Lambda {\it N}$ coupling may affect
not only the magnitude of the splittings but also the ordering of the levels
for the spin doublets.
In fact, the inversion of levels appears, in Fig.~\ref{fig:3},
as seen in the results of the first-excited doublet ($1_{2}^{-},2_{1}^{-}$)
for the NSC97f
at $\rho _{\Sigma}=1$ and $2$ though the splitting energies are very small.

Tzeng {\it et al}. have shown that the $1^{-}$ states for both
of the doublets become more attractive compared to the other spin partners
if they take into account the effect of the $\Lambda$-$\Sigma$
coupling three-body force~\cite{Tzeng00}.
In our approach, the effect of the $\Lambda$-$\Sigma$ three-body force
is automatically taken into account when we include the $\Sigma {\it N}$
channel in the model space, as discussed before.
In our results the trend of the $\Lambda$-$\Sigma$ three-body effect
on the first-excited doublet agrees with the results
by them, but that on the ground-state doublet does not necessarily
agree for the NSC97f.

We should say, however, that
the $\Lambda$-$\Sigma$ three-body effect on the energy levels may appear
more clearly if we use a ${\it YN}$ potential which
has a strong $\Sigma {\it N}$-$\Lambda {\it N}$ interaction such as
the NSC89 potential.
As a matter of fact, both splittings of the doublets become larger
as $\rho _{\Sigma}$ increases, if we use the NSC89.
For example, the splitting energies of the ground-state doublet
are $0.17$MeV and $0.61$MeV, respectively,
for the cases of $``\Lambda {\it N}"$ and $ \rho_{\Sigma}=5$,
and those of the first-excited doublet, $0.26$MeV and $0.81$MeV.
We note here that both of the $1^{-}$ states are always more attractive
than the other spin partners for the NSC89 regardless of the values of
$\rho _{\Sigma}$,
which is a different feature from the results for the NSC97 models.
Our results of the two doublets for the NSC89
agree fairly well with their results~\cite{Tzeng00}.

In Table~\ref{tab:table2}, the calculated energies of the low-lying states
for $\rho _{\Sigma}=5$ are tabulated for the NSC97a-f potentials.
The results for ``$\Lambda {\it N}$" are also shown
in parentheses for reference.
Smooth variation of the splittings of the ground and first-excited doublets
with negative parity is observed when we change the ${\it YN}$ potentials
from the NSC97a to NSC97f. 
It may be considered that this variation is a reflection of the different
strengths of the spin-dependent interactions originated from the variation of
the magnetic $F/(F+D)$ ratio $\alpha _{V}^{m}$ for the vector mesons in the
NSC97a-f~\cite{NSC97}.
The gradual change can be also seen in the $\Lambda$ spin-orbit
splitting in $_{\Lambda}^{17}$O as shown in Table~\ref{tab:table1}.
We note that the $1_{1}^{-}$ state of the ground-state doublet lies
in energy above the $0_{1}^{-}$ state for
the NSC97a-f.

\begin{figure}[b]
\includegraphics[height=.42\textheight]{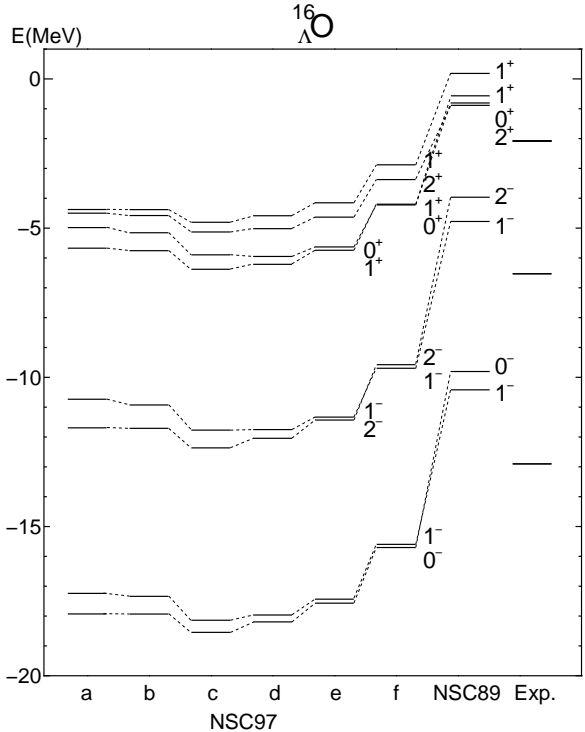}
\caption{\label{fig:4} Energy levels
in $_{\Lambda}^{16}$O.
The calculated results were obtained for $\rho _{\Sigma}=5$.
The experimental levels were taken from Ref.~\cite{Tamura94}.}
\end{figure}

In the E930 experiment at BNL, two $\gamma$-transitions from the
$1_{2}^{-}$ state to the ground-state doublet ($0_{1}^{-}$, $1_{1}^{-}$)
are expected to be observed~\cite{Tamura01}.
From these the magnitude of the splitting of the ground-state doublet
should be established.
Our results might help to constrain parameters
such as $\alpha _{V}^{m}$ for the NSC97 models
in determining ${\it YN}$ interactions.

As for the positive parity states,
one can see that the calculated results have very weak dependence on
$\rho _{\Sigma}$ for the NSC97d and NSC97f in Fig.~\ref{fig:3}.
It has been confirmed that
similar tendency is observed for the other NSC97 models.
In Table~\ref{tab:table2}, the calculated energies of the positive- and
negative-parity states have been tabulated
for the NSC97a-f potentials.
In Fig.~\ref{fig:4}, we also show the calculated energy levels using
the NSC97a-f and NSC89 potentials for $\rho _{\Sigma}=5$
together with the experimental levels.

An interesting feature can be seen
concerning the relative position of the $0^{+}_{1}$ and $2^{+}_{1}$ states.
Our results of the $0^{+}_{1}$ state are below the $2^{+}_{1}$ state
in energy, except for the NSC89.
The relative positions of the $0^{+}_{1}$ and $2^{+}_{1}$ states in our results
show a different feature
from the results of the shell-model calculations
by other groups~\cite{Motoba98,Tzeng00}
though the same ${\it YN}$ interactions are employed.
In those calculations, the $2^{+}_{1}$ state lies below the $0^{+}_{1}$ state.
In a simple picture, the $0_{1}^{+}$ and $2_{1}^{+}$ states have
the main components composed of 
$[0p_{1/2}^{\Lambda}, 0p_{1/2}^{-1}]$ and 
$[0p_{3/2}^{\Lambda}, 0p_{1/2}^{-1}]$, respectively.
Thus, the $2^{+}_{1}$ state should be below the $0^{+}_{1}$ state
in connection with the positions of the $0p_{3/2}^{\Lambda}$ and
$0p_{1/2}^{\Lambda}$ states which are separated in energy by
the $\Lambda$ spin-orbit splitting in $_{\Lambda}^{17}$O.

We found that the inversion of the levels in our results was caused mainly by
the parity-mixing intershell coupling in $1\hbar \omega$ excitation
as discussed by Motoba~\cite{Motoba98}.
As a unique feature of the structure of $\Lambda$ hypernuclei,
negative- and positive-parity nuclear core states can couple
in the same energy region through a transition of $\Lambda$ states such as
the $0p_{1/2}^{\Lambda}$ state to the $0s_{1/2}^{\Lambda}$ state.
In other words, even if the $1p$-$1h$ excitation of
nuclear core pushes its energy up by about $1\hbar \omega$,
the energy can be compensated by the transition of $\Lambda$ states
in $1\hbar \omega$ energy region.
In our calculation using the NSC97f and the Paris potentials,
the $0_{1}^{+}$ state does not have
the simple $[0p^{\Lambda}_{1/2},0p^{-1}_{1/2}]$ configuration,
but a rather complex structure, as illustrated
in Table~\ref{tab:table3}.
We see that the probability for the last configuration
which includes the positive-parity core-excited state is the
same order of magnitude as that for the first configuration
which includes the negative-parity state of the core nucleus.

%------percentage analysis of mixing amplitudes-----------------
\begin{table}[b]
\caption{\label{tab:table3} The percentage analysis of the low-lying positive
parity states for $\rho _{\Sigma}=5$.
The percentage for each configuration denotes
the probability, namely, the square of mixing amplitude.
The values in parentheses denote the results using
the renormalized ``$\Lambda {\it N}$" effective interaction.
The NSC97f and the Paris potentials are employed for the ${\it YN}$ and
${\it NN}$ interactions, respectively.}
\begin{ruledtabular}
    \begin{tabular}{cccc}
   Configuration                      & $0_{1}^{+}$ & $1_{1}^{+}$ & $2_{1}^{+}$\\ \hline
$[0p^{\Lambda}_{1/2}, 0p^{-1}_{1/2}]$ & $30.1$\%    & $ 1.9$\%    & $   0$\%  \\
                                      &($27.4$\%)   &($ 1.8$\%)   &($   0$\%) \\
$[0p^{\Lambda}_{3/2}, 0p^{-1}_{1/2}]$ & $   0$\%    & $37.8$\%    & $65.6$\%  \\
                                      &($   0$\%)   &($34.3$\%)   &($64.0$\%) \\
$[0s^{\Lambda}_{1/2}, 0s^{-1}_{1/2}]$ & $12.1$\%    & $ 9.0$\%    & $   0$\%  \\
                                      &($12.3$\%)   &($ 9.5$\%)   &($   0$\%) \\
$[0s^{\Lambda}_{1/2}, 0d_{5/2},0p^{-1}_{3/2},0p^{-1}_{1/2}]$
                                      & $33.0$\%    & $26.6$\%    & $ 6.9$\%  \\
                                      &($35.8$\%)   &($30.2$\%)   &($ 8.4$\%) \\
    \end{tabular}
\end{ruledtabular}
\end{table}
%-------------------------------------------

As for the $2_{1}^{+}$ state, however, such a strong effect of
the parity-mixing intershell coupling does not appear.
The dominant configuration is only
$[0p^{\Lambda}_{3/2},0p^{-1}_{1/2}]$
which is the natural configuration with the lowest unperturbed energy.
It should be noted that the $2^{+}_{1}$ state can not be constructed
from the configurations including the $0s^{\Lambda}_{1/2}$ state
in the space of the $1\Lambda$-$1h$ configuration.
The $[0s^{\Lambda}_{1/2},0s^{-1}_{1/2}]$ configuration
can couple with the $p$-$h$ excited configuration
$[0s^{\Lambda}_{1/2},0d_{5/2},0p^{-1}_{3/2},0p^{-1}_{1/2}]$
through the ${\it NN}$ effective interaction
to construct the $0^{+}_{1}$ state.
Thus, the energies of the $0_{1}^{+}$ and $2_{1}^{+}$ states 
are dependent on the adopted ${\it NN}$ interaction 
as well as the ${\it YN}$ interaction.
The same discussion on the parity-mixing intershell coupling as the $0_{1}^{+}$
state applies to the $1_{1}^{+}$ state
as we see from Table~\ref{tab:table3}.
Although the parity-mixing intershell coupling also affects
the $2_{1}^{+}$ state,
the effect is considerably smaller than the $0_{1}^{+}$ and $1_{1}^{+}$ states.
Therefore, we conclude that the parity-mixing intershell coupling
strongly affects special states such as the $0_{1}^{+}$ and $1_{1}^{+}$
states with the help of the ${\it NN}$ effective interaction.

Concerning the comparison with the experimental levels,
we may say that the calculated results of the excitation spectra from
the ground state agree well with the experimental values, on the whole,
as shown in Fig.~\ref{fig:4}.
The relative energy between the lowest two levels
in the experimental data corresponds to the spin-orbit splitting energy
of the nucleon.
Our results of the splitting between the lowest two bunched levels show a good
agreement with the corresponding experimental values.
These splittings of the calculated results are obtained,
reflecting the property of the adopted ${\it NN}$ interaction
which is the Paris potential in the present study.

In the present shell-model calculations,
we do not employ the experimental single-particle energies
of $\Lambda$, $\Sigma$, and nucleons.
The results thus obtained show directly the differences in properties between
the ${\it YN}$ interactions.
In general, the effective ${\it YN}$ and ${\it NN}$ interactions are derived
dependently on the single-particle energies
of $\Lambda$, $\Sigma$, and nucleons.
These single-particle energies are determined by both of the spin-dependent
and spin-independent interactions.
In this context, our results reflect not only the spin-dependent interaction
but also the spin-independent one of
the free ${\it YN}$ and ${\it NN}$ interactions.

We here make some comments on the results
obtained by Tzeng {\it et al}.~\cite{Tzeng00,Tzeng02}.
It seems that our results of $_{\Lambda}^{16}$O considerably differ
from their results at first sight.
As a matter of fact, the dependence of the calculated energy levels
in the absolute value on the ${\it YN}$ interactions is
different from each other.
This is mainly because of the difference of the treatment of
the single-particle energies.
In their method, a common set of the semi-empirical single-particle energies is
employed in the calculations using various ${\it YN}$ interactions.
Therefore, there are considerable differences in
the absolute values of the energy levels between their and our results.
However, as far as we are concerned with the excitation spectra,
especially the splittings of the two doublets with negative parity,
our results of $_{\Lambda}^{16}$O for the NSC97a-f and NSC89
potentials are consistent, on the whole, with the results obtained by them.
As a general feature, our results of the splittings of the two doublets
show a little smaller values than their results.

We move to the discussion on the dependence of the calculated results
on the ${\it YN}$ interactions.
In our results of the ground-state doublet in $_{\Lambda}^{16}$O,
the result for the NSC97c is the most attractive in the NSC97 models,
and that for the NSC97f is the least attractive.
One may consider that this tendency is not consistent with results of
$_{\Lambda}^{4}$He in a recent calculation~\cite{Nogga02}.
In that study, the binding energy of the ground $0^{+}$ state
for the NSC97f is the most attractive, and the result for the NSC97d
is the least attractive.
Results for the NSC97a-c potentials have not been given in their paper.
This question of the inconsistency can be solved by analyzing matrix elements
of our $\Lambda {\it N}$ effective interaction.

\begin{figure}[b]
\includegraphics[height=.31\textheight]{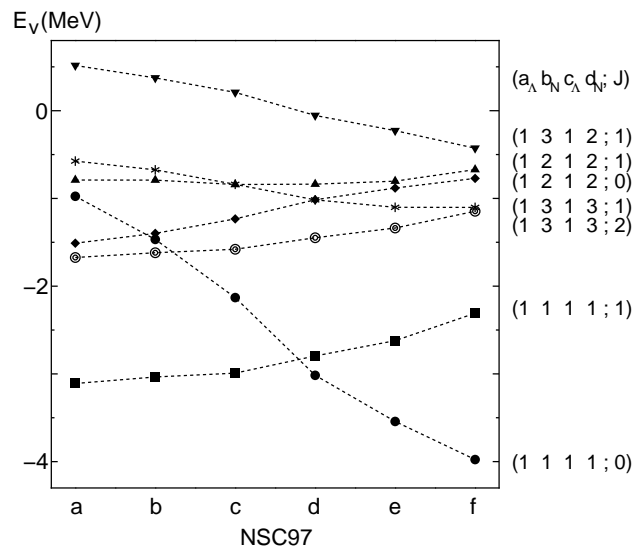}
\caption{\label{fig:5} Dependence of some of the representative matrix elements
of the renormalized ``$\Lambda {\it N}$" effective interaction
$\langle a_{\Lambda}b_{\it N}|\tilde{v}_{\Lambda {\it N}}
|c_{\Lambda}d_{\it N} \rangle _{J,T=1/2}$ on the NSC97 models
in $_{\Lambda}^{16}$O.
The single-particle states are labeled as $1=0s_{1/2}$, $2=0p_{1/2}$, and
$3=0p_{3/2}$.}
\end{figure}

In Fig.~\ref{fig:5}, some of the representative matrix elements of
the renormalized ``$\Lambda {\it N}$" effective interaction
for the NSC97 models are shown.
We see that the matrix elements vary almost linearly
from the NSC97a to NSC97f.
In the shell-model language, the dominant contribution of the matrix element
to the ground $0^{+}$ state in $_{\Lambda}^{4}$He should be
$\langle 0s_{1/2}0s_{1/2}|\tilde{v}_{\Lambda {\it N}}|
0s_{1/2}0s_{1/2}\rangle _{J=0}$, namely, the spin-singlet $s$-wave interaction.
This matrix element for the NSC97f is the most attractive
in the NSC97 models as seen in Fig.~\ref{fig:5},
and the dependence on the ${\it YN}$ interactions is consistent with
the results of $_{\Lambda}^{4}$He.
On the other hand, the matrix element
$\langle 0s_{1/2}0s_{1/2}|\tilde{v}_{\Lambda {\it N}}|
0s_{1/2}0s_{1/2}\rangle _{J=1}$
which represents the spin-triplet $s$-wave interaction contributes dominantly
to the $1^{+}$ state, namely, the spin partner
of the $0^{+}$ state in $_{\Lambda}^{4}$He.
The value of this matrix element for the NSC97f is the least attractive
in the NSC97 models.
This trend is observed in their results of the $1^{+}$ state
in $_{\Lambda}^{4}$He~\cite{Nogga02}.
Therefore, our results are not inconsistent with
the results of $_{\Lambda}^{4}$He.

In the calculation of the ground-state doublet with negative parity
in $_{\Lambda}^{16}$O,
the other matrix elements in Fig.~\ref{fig:5} are important.
These matrix elements include effects of the $p$-wave interaction
in addition to the $s$-wave one.
The $p$-wave interaction also contributes to the $\Lambda$ single-particle
energy for the $0s_{1/2}$ state,
and thus the structure of $_{\Lambda}^{16}$O becomes more complex than
$_{\Lambda}^{4}$He.
As a result, the energies for the NSC97c are the most attractive
in the NSC97 models in $_{\Lambda}^{16}$O.
It may be considered that the study of $_{\Lambda}^{16}$O is useful 
to investigate properties of higher partial-wave interactions
such as the $p$-wave interaction.

Finally, we move back to the discussion on the comparison of our results and
the experimental values.
Roughly speaking, the experimental energy levels lie
between the two results for the NSC97f and NSC89 potentials.
As we mentioned before, however, effects of the many-body effective interaction
are not included in the present calculation.
In the previous study, the effect of the three-body cluster (TBC) terms on
the $\Lambda$ single-particle energy in $_{\Lambda}^{17}$O was
investigated~\cite{Fujii00}.
It was confirmed that the TBC terms caused
a repulsive contribution of about $1$MeV to the $\Lambda$ single-particle
energy for the $0s_{1/2}$ state in $_{\Lambda}^{17}$O for the NSC97f.
This suggests that the TBC effect shifts the energy levels of
the negative-parity states in $_{\Lambda}^{16}$O
repulsively to the same extent,
if we take into account this effect in the present study. 
Thus, the NSC97f potential might be favorable in the NSC97 models
for the calculation for $_{\Lambda}^{16}$O.

\section{\label{sec:Remarks}Concluding Remarks}

Shell-model calculations for $_{\Lambda}^{17}$O and $_{\Lambda}^{16}$O in
a large model space have been performed.
By introducing a new model space including $\Sigma {\it N}$ states,
we have calculated effective interactions which include the
$\Sigma {\it N}$-$\Lambda {\it N}$ coupling terms. 
As far as we know, the degrees of freedom of $\Sigma$
in addition to $\Lambda$ and nucleons
have been explicitly introduced in the shell-model calculations
for the first time.
The effective interactions and the single-particle energies employed in the
shell-model calculations have been microscopically
derived from the NSC97a-f and NSC89 ${\it YN}$ interactions and
the Paris ${\it NN}$ interaction within the framework of the UMOA.

It has been confirmed that the results of the present shell-model
diagonalization for $_{\Lambda}^{17}$O with the NSC97a-f potentials
agree well with those of our previous study in which the calculation was
performed perturbatively.
The $\Lambda$ spin-orbit splitting energies obtained are
$0.44$MeV to $0.95$MeV for the NSC97a to NSC97f.
These values seem to be considerably larger than the value
suggested from the experimental result of $_{\Lambda}^{13}$C
which has been established recently.

It has been found that a drastic change in the structure of $_{\Lambda}^{16}$O
induced by the $\Sigma$ degrees of freedom does not occur
as far as we employ the NSC97a-f potentials.
The $\Sigma$ degrees of freedom give rise to a small effect on
the first-excited doublet ($1_{2}^{-}$, $2_{1}^{-}$) in $_{\Lambda}^{16}$O.
However, if we use the NSC89 potential
which has a strong $\Sigma {\it N}$-$\Lambda {\it N}$ interaction,
the splittings of the ground and first-excited doublets, respectively,
($0_{1}^{-}$, $1_{1}^{-}$) and ($1_{2}^{-}$, $2_{1}^{-}$) are enlarged.
The magnitude of the splitting of the ground-state doublet
gradually decreases as $0.69$MeV to $0.11$MeV from the NSC97a to NSC97f.
We should note that the $0_{1}^{-}$ state lies below the $1_{1}^{-}$ state
in energy for the NSC97 models.
On the other hand, the $1_{1}^{-}$ state is below the $0_{1}^{-}$ state
for the NSC89.
In the E930 experiment at BNL, the magnitude of
the ground-state doublet should be determined in the near future,
which would give useful information on the underlying properties
of the ${\it YN}$ interaction.

Effects of the parity-mixing intershell coupling on $1\hbar \omega$
excited states have been investigated.
It has been found that the parity-mixing intershell coupling plays
an important role in the structure of
the $0_{1}^{+}$ and $1_{1}^{+}$ states in $_{\Lambda}^{16}$O
with the help of the ${\it NN}$ effective interaction.
As a result, the $0_{1}^{+}$ and $1_{1}^{+}$ states have complex structures.
On the other hand,
the parity-mixing intershell coupling on the $2_{1}^{+}$ state is
less active than the $0_{1}^{+}$ and $1_{1}^{+}$ states.

In conclusion, the present shell-model results,
especially, the excitation spectra
have shown a good agreement with the experimental levels
on the whole,
even though our calculation method is fully microscopic and does not
include any experimental values and adjustable parameters.
The experimental levels are between the two results for the NSC97f and NSC89.
In the near future, some fine structures of $\Lambda$ hypernuclei
reflecting the properties of the underlying ${\it YN}$ interaction
would be revealed experimentally.
We hope that our method will help to bridge between the ${\it YN}$ interaction
and the $\Lambda$ hypernuclear structure microscopically,
and give useful constraint to determine ${\it YN}$ interactions
more realistically.

\begin{acknowledgments}
The authors are grateful to Yiharn Tzeng for the calculations for
$_{\Lambda}^{16}$O using the NSC97a-e potentials in reply to our request.
One of the authors (S.~F.) would like to thank H. Tamura
for useful discussion about the E930 experiment at BNL.
He also acknowledges the Special Postdoctoral Researchers Program of RIKEN.
\end{acknowledgments}

\appendix
\section{Matrix Elements of $\Lambda {\it N}$ Effective Interactions}
In Table~\ref{tab:table4}, we tabulate representative matrix elements of
the renormalized ``$\Lambda {\it N}$" effective interactions for the NSC97a-f
potentials for reference.
The ``$\Lambda {\it N}$" means that the $\Sigma {\it N}$ channel
is not included in the model space in determining
the ``$\Lambda {\it N}$" effective interactions.
The results using these potentials are tabulated in parentheses
in Tables~\ref{tab:table1}
and~\ref{tab:table2}, and shown in Figs.~\ref{fig:2} and~\ref{fig:3}.
The matrix elements in the $0s$ and $0p$ shells are given.

\begin{table*}[t]
\caption{\label{tab:table4}
Matrix elements of the
renormalized ``$\Lambda {\it N}$" effective interaction
$\langle a_{\Lambda}b_{\it N}|\tilde{v}_{\Lambda {\it N}}
|c_{\Lambda}d_{\it N} \rangle _{J,T=1/2}$
for the NSC97a-f potentials.
The single-particle states are labeled as $1=0s_{1/2}$, $2=0p_{1/2}$, and
$3=0p_{3/2}$. All energies are in MeV.}
\begin{ruledtabular}
    \begin{tabular}{cccccrrrrrr}
$a_{\Lambda}$ & $b_{\it N}$ & $c_{\Lambda}$ & $d_{\it N}$ & $J$ &
  NSC97a& NSC97b & NSC97c & NSC97d & NSC97e & NSC97f \\ \hline
 $1$   &  $1$ &  $1$   &  $1$ & $0$ & $-0.975$ & $-1.467$ & $-2.130$ & $-3.016$ & $-3.542$ & $-3.976$ \\
 $1$   &  $1$ &  $1$   &  $1$ & $1$ & $-3.109$ & $-3.035$ & $-2.991$ & $-2.795$ & $-2.621$ & $-2.308$ \\
 $1$   &  $2$ &  $1$   &  $2$ & $0$ & $-1.510$ & $-1.398$ & $-1.233$ & $-1.015$ & $-0.881$ & $-0.771$ \\
 $1$   &  $2$ &  $1$   &  $2$ & $1$ & $-0.790$ & $-0.791$ & $-0.841$ & $-0.837$ & $-0.803$ & $-0.671$ \\
 $1$   &  $3$ &  $1$   &  $2$ & $1$ & $ 0.516$ & $ 0.374$ & $ 0.209$ & $-0.054$ & $-0.227$ & $-0.427$ \\
 $1$   &  $3$ &  $1$   &  $3$ & $1$ & $-0.574$ & $-0.674$ & $-0.839$ & $-1.016$ & $-1.101$ & $-1.102$ \\
 $1$   &  $3$ &  $1$   &  $3$ & $2$ & $-1.673$ & $-1.619$ & $-1.579$ & $-1.448$ & $-1.337$ & $-1.146$ \\
 $2$   &  $1$ &  $1$   &  $2$ & $0$ & $-1.583$ & $-1.617$ & $-1.727$ & $-1.750$ & $-1.713$ & $-1.538$ \\
 $2$   &  $1$ &  $1$   &  $2$ & $1$ & $ 0.990$ & $ 0.917$ & $ 0.817$ & $ 0.658$ & $ 0.554$ & $ 0.449$ \\
 $2$   &  $1$ &  $1$   &  $3$ & $1$ & $-1.170$ & $-1.306$ & $-1.477$ & $-1.702$ & $-1.832$ & $-1.950$ \\
 $2$   &  $2$ &  $1$   &  $1$ & $0$ & $-0.083$ & $-0.274$ & $-0.536$ & $-0.883$ & $-1.088$ & $-1.247$ \\
 $2$   &  $2$ &  $1$   &  $1$ & $1$ & $ 0.301$ & $ 0.260$ & $ 0.261$ & $ 0.202$ & $ 0.133$ & $-0.019$ \\
 $2$   &  $3$ &  $1$   &  $1$ & $1$ & $-1.087$ & $-1.052$ & $-1.018$ & $-0.931$ & $-0.865$ & $-0.761$ \\
 $2$   &  $1$ &  $2$   &  $1$ & $0$ & $-1.783$ & $-1.677$ & $-1.531$ & $-1.317$ & $-1.177$ & $-1.037$ \\
 $2$   &  $1$ &  $2$   &  $1$ & $1$ & $-1.170$ & $-1.188$ & $-1.258$ & $-1.277$ & $-1.255$ & $-1.129$ \\
 $2$   &  $2$ &  $2$   &  $2$ & $0$ & $ 0.243$ & $ 0.166$ & $ 0.013$ & $-0.140$ & $-0.211$ & $-0.178$ \\
 $2$   &  $2$ &  $2$   &  $2$ & $1$ & $-1.271$ & $-1.222$ & $-1.200$ & $-1.098$ & $-1.002$ & $-0.823$ \\
 $2$   &  $3$ &  $2$   &  $2$ & $1$ & $-0.208$ & $-0.223$ & $-0.227$ & $-0.243$ & $-0.260$ & $-0.300$ \\
 $2$   &  $3$ &  $2$   &  $3$ & $1$ & $-1.040$ & $-0.961$ & $-0.872$ & $-0.715$ & $-0.605$ & $-0.452$ \\
 $2$   &  $3$ &  $2$   &  $3$ & $2$ & $-1.142$ & $-1.201$ & $-1.294$ & $-1.382$ & $-1.419$ & $-1.406$ \\
 $3$   &  $1$ &  $1$   &  $2$ & $1$ & $ 1.298$ & $ 1.434$ & $ 1.603$ & $ 1.823$ & $ 1.950$ & $ 2.062$ \\
 $3$   &  $1$ &  $1$   &  $3$ & $1$ & $-0.118$ & $ 0.051$ & $ 0.272$ & $ 0.588$ & $ 0.784$ & $ 0.969$ \\
 $3$   &  $1$ &  $1$   &  $3$ & $2$ & $ 1.433$ & $ 1.414$ & $ 1.410$ & $ 1.353$ & $ 1.295$ & $ 1.195$ \\
 $3$   &  $2$ &  $1$   &  $1$ & $1$ & $ 1.087$ & $ 1.052$ & $ 1.018$ & $ 0.931$ & $ 0.865$ & $ 0.761$ \\
 $3$   &  $3$ &  $1$   &  $1$ & $0$ & $-0.117$ & $-0.388$ & $-0.757$ & $-1.249$ & $-1.538$ & $-1.764$ \\
 $3$   &  $3$ &  $1$   &  $1$ & $1$ & $-1.163$ & $-1.105$ & $-1.075$ & $-0.961$ & $-0.858$ & $-0.669$ \\
 $3$   &  $1$ &  $2$   &  $1$ & $1$ & $-0.563$ & $-0.402$ & $-0.211$ & $ 0.089$ & $ 0.286$ & $ 0.507$ \\
 $3$   &  $2$ &  $2$   &  $2$ & $1$ & $ 0.273$ & $ 0.287$ & $ 0.283$ & $ 0.297$ & $ 0.314$ & $ 0.363$ \\
 $3$   &  $2$ &  $2$   &  $3$ & $1$ & $ 0.947$ & $ 0.956$ & $ 0.986$ & $ 0.982$ & $ 0.961$ & $ 0.903$ \\
 $3$   &  $2$ &  $2$   &  $3$ & $2$ & $-0.659$ & $-0.569$ & $-0.457$ & $-0.283$ & $-0.169$ & $-0.050$ \\
 $3$   &  $3$ &  $2$   &  $2$ & $0$ & $-0.371$ & $-0.648$ & $-0.986$ & $-1.471$ & $-1.771$ & $-2.070$ \\
 $3$   &  $3$ &  $2$   &  $2$ & $1$ & $ 0.933$ & $ 0.914$ & $ 0.919$ & $ 0.886$ & $ 0.843$ & $ 0.754$ \\
 $3$   &  $3$ &  $2$   &  $3$ & $1$ & $-0.934$ & $-0.913$ & $-0.898$ & $-0.848$ & $-0.805$ & $-0.735$ \\
 $3$   &  $3$ &  $2$   &  $3$ & $2$ & $ 0.145$ & $ 0.245$ & $ 0.371$ & $ 0.551$ & $ 0.663$ & $ 0.780$ \\
 $3$   &  $1$ &  $3$   &  $1$ & $1$ & $-0.646$ & $-0.779$ & $-0.986$ & $-1.221$ & $-1.342$ & $-1.378$ \\
 $3$   &  $1$ &  $3$   &  $1$ & $2$ & $-1.920$ & $-1.863$ & $-1.822$ & $-1.681$ & $-1.561$ & $-1.352$ \\
 $3$   &  $2$ &  $3$   &  $2$ & $1$ & $-0.950$ & $-0.865$ & $-0.764$ & $-0.597$ & $-0.486$ & $-0.344$ \\
 $3$   &  $2$ &  $3$   &  $2$ & $2$ & $-1.084$ & $-1.138$ & $-1.228$ & $-1.310$ & $-1.343$ & $-1.324$ \\
 $3$   &  $3$ &  $3$   &  $2$ & $1$ & $ 0.974$ & $ 0.949$ & $ 0.932$ & $ 0.878$ & $ 0.829$ & $ 0.749$ \\
 $3$   &  $3$ &  $3$   &  $2$ & $2$ & $-0.171$ & $-0.275$ & $-0.402$ & $-0.586$ & $-0.702$ & $-0.824$ \\
 $3$   &  $3$ &  $3$   &  $3$ & $0$ & $-0.043$ & $-0.315$ & $-0.707$ & $-1.203$ & $-1.485$ & $-1.662$ \\
 $3$   &  $3$ &  $3$   &  $3$ & $1$ & $-0.923$ & $-0.849$ & $-0.809$ & $-0.670$ & $-0.549$ & $-0.323$ \\
 $3$   &  $3$ &  $3$   &  $3$ & $2$ & $-0.308$ & $-0.352$ & $-0.430$ & $-0.505$ & $-0.540$ & $-0.523$ \\
 $3$   &  $3$ &  $3$   &  $3$ & $3$ & $-1.738$ & $-1.703$ & $-1.686$ & $-1.597$ & $-1.514$ & $-1.368$ \\
    \end{tabular}
\end{ruledtabular}
\end{table*}

%\bibliography{o16l}% Produces the bibliography via BibTeX.

\end{document}